# Measuring the influence of a journal using impact and diffusion factors

**S.A. Sanni and A.N. Zainab**

Digital Library Research Group, Faculty of Computer Science and Information Technology,
University of Malaya, Kuala Lumpur, MALAYSIA
e-mail: sanni_shams@yahoo.com, zainab@um.edu.my

## ABSTRACT

*Presents the result of the calculated ISI equivalent Impact Factor, Relative Diffusion Factor (RDF), and Journal Diffusion Factor (JDF) for articles published in the Medical Journal of Malaysia (MJM) between the years 2004 and 2008 in both their synchronous and diachronous versions. The publication data are collected from MyAis (Malaysian Abstracting & Indexing system) while the citation data are collected from Google Scholar. The values of the synchronous JDF ranges from 0.057 – 0.14 while the diachronous JDF ranges from 0.46 – 1.98. The high diachronous JDF is explained by a relatively high number of different citing journals against the number of publications. This implies that the results of diachronous JDF is influenced by the numbers of publications and a good comparison may be one of which the subject of analysis have similar number of publications and citations period. The yearly values of the synchronous RDF vary in the range of 0.66 – 1.00 while diachronous RDF ranges from 0.62 – 0.88. The result shows that diachronous RDF is negatively correlated with the number of citations, resulting in a low RDF value for highly cited publication years. What this implies in practice is that the diffusion factors can be calculated for every additional year at any journal level of analysis. This study demonstrates that these indicators are valuable tools that help to show development of journals as it changes through time.*

## INTRODUCTION

Various indicators have been introduced to evaluate journals and the most popular and widely used being the ISI derived journal impact factor (JIF). The IF is probably the most used metric to judge the influence of a scientific journal, and for gauging the visibility of journals on the research front (Rousseau 2002; Sharma 2007; Della Sala and Grafman 2009; Franceschet 2010). However, a number of studies have highlighted the limitations of JIF as the sole indicator of a journal's influence because it failures to measure the breadth of influence across the literature of a particular journal title (Rowland 2002; Frandsen, Rousseau and Rowlands 2006). The obvious shortcoming of the JIF is its limiting coverage to a 2 – year window which may favor some journals especially older ones and penalize newer titles (Della Sala and Grafman 2009; Swartz 2009). Also, the number of citations an article receives will only accumulate over a number of years, thus, Rousseau (2002), Archambault and Gagné (2004), Fok and Franses (2007) and Haddow and Genoni (2009) suggest that longer citation periods should be taking into account.

In order to supplement and complement the JIF and to further understand the spread and breadth of a journal's influence or shifts in scholarly research as they change over time, other form of indicators have been developed. Examples are: cited half-life, journal diffusion factors (JDF), relative diffusion factors (RDF), popularity factor, h-Index, journal downloads and immediacy index.

In this paper, we present the results of calculated ISI equivalent Journal Impact Factor (JIF), Relative Diffusion Factor (RDF), and Journal Diffusion Factor (JDF) for articles published in the *Medical Journal of Malaysia (MJM)* between the years 2004 and 2008 in both their synchronous and diachronous versions. The publication data were collected from *MyAis* (Malaysian Abstracting & Indexing System http://myais.fsktm.um.edu.my/) while the citation data were collected from *Google Scholar*, sourced from the Publish or Perish software (http://wwwxliarzing.com/resources.htm).

## RELATED LITERATURE

Ingwersen et al. (2001) gave an overview of the main data of a publication-citation matrix and showed how impact factors were defined, and, in particular, pointed out the difference between the synchronous and the diachronous impact factor. It was shown that the diachronous impact factor compares like with like and this is preferred in evaluation studies. The diachronous impact factor can also be applied to a larger group of publications, including conference proceedings. Furthermore, problems associated with the use of impact factors, synchronous or diachronous, for research evaluation were discussed and recommended that diachronous impact factors be used as a measure of the expected impact of the articles in journals.

Rowlands (2002) argued that the bibliometric indicators such as the JIF, immediacy index, and cited half-life have commonly used for measuring the quality of research but offered little insight into the trans-disciplinary reception of journals. He introduced a new bibliometric measure called *Journal Diffusion Factor* (JDF). To communicate this concept, Rowlands used the "pebble – pond metaphor". He pointed out that new ideas can be viewed like pebbles thrown into a pond and the surface of the pond represent the general research literature. This situation provided two potentially useful metaphor; the size of the splash as the pebble hits the surface of the water, and the characteristics of the resulting ripples. He noted that JIF only measured the mass of the pebble and in order to understand the splash they should be read carefully alongside immediacy and cited half-life, and to understand the effect of subsequent ripples, the "breadth" of the reception of a particular journal or JDF should be utilized. The author defined the JDF as the average number of citing journals per 100 source citations within a given time window. He concluded that JDF is different from JIF, both conceptually and in terms of its statistical behavior, to be seriously considered in research or journal evaluations.

$$\text{Rowland's JDF} = JDF\,(n_p, n_c, y_p, y_c, j) = \frac{R\,(k, i, j) * 100}{\sum_{i=y_p}^{y_p+n_p-1} \sum_{k=y_c}^{y_c+n_c-1} Cy\,(k, i, j)}$$

*np*        equals the length of the publication period measured in years,
*nc*        equals the length of the citation window measured in years,
*yp*        is the beginning year of the publication period,
*yc*        is the beginning year of the citation window,
*i*         is the publication year(s),
*k*         is the citation year(s),
*j*         is the cited journal under investigation,
*Cy(k,i,j)* is the number of citations that the documents published in year(s) *k*
            of the journal *j* receives in the year(s) *i*.
*R(k,i,j)*  is the number of different journals that cites the documents
            published in year(s) *k* of the journal *j* in the year(s) *i*.

Frandsen et al. (2004), Egghe (2005) and Frandsen, Rousseau and Rowlands (2006) presented new approaches to measuring research influence. The aim was to complement existing metrics with a series of formally described diffusion factors. Using a publication-citation matrix as an organizing construct, the author's developed formal descriptions of two forms of diffusion metric: "*relative diffusion factors*" and "*journal diffusion* factors" in both their synchronous and diachronous forms. Results from worked examples showed that, Diffusion factors captured different aspects of the citation reception process than existing bibliometric measures and that they can be applied at the whole journal level or for sets of articles. It is also possible to provide a richer evidence base for citation analyses than the traditional measure alone. This paper will also test these metrics with data samples of from *MJM using Frandsen's* formula.

$$\text{Frandsen's JDF} = JDF\,(n_p,\, n_c,\, y_p,\, y_c,\, j) = \frac{R\,(k,\, i,\, j)}{\displaystyle\sum_{t=yp}^{yp\,+\,np\,-\,1} Py(k,j)}$$

*np*        equals the length of the publication period measured in years,
*nc*        equals the length of the citation window measured in years,
*yp*         is the beginning year of the publication period,
*yc*        is the beginning year of the citation window,
*i*          is the publication year(s),
*k*          is the citation year(s),
*j*         is the cited journal under investigation,
*Py(k,j)*   is the number of citable units published in year(s) *k* of the journal *j*.
*R(k,i,j)*  is the number of different journals that cites the documents published in year(s)*k* of
            the journal *j* in the year(s) *i*.

Other indicators has also been proposed, such as the Journal Citation Image (JCI) by Bonnevie-Nebelong (2006) who did an analysis on *Journal of Documentation* (*JDOC*) which, is then compared to JASIS(T) and the *Journal of Information Science* (JIS). The JCI is based on Frandsen's (2004) new JDF and Journal Co-Citation Analysis. The JDF revealed that *JDOC* reaches far out into the scientific community and there was a tendency positive correlation between increasing JDF and JIF. The overall results gave multifaceted pictures of *JDOC* and indicated the versatility of JDF and a measure to supplement other indicators such as the JIF.

Fok and Franses (2007) presented an extended version of the diffusion model by considering the relationships between key characteristics of the diffusion process and features of the articles. More specifically, parameters measuring citations' ceiling and the timing of peak citations are correlated with specific features of the articles like the number of pages and the number of authors. The model was illustrated for citations to articles that were published in *Econometrica* and the *Journal of Econometrics*. The first consequence of the analysis is that there might be a need to re-consider the current practices used by the SSCI, which ranked journals according to citations within 2 years after publication, because the number of citations actually varies over time. In addition, it is likely that journals vary with respect to the citation diffusion of their articles, due to the possibility that each journal might have a different type of audience with a different citation style. They concluded that the new model allows for a compact description of the citation process, and make possible comparisons across disciplines.

Haddow (2008) presented the reports of a pilot study undertaken to test Roland's JDF (Roland 2002) as an alternative to JIFs for ranking journals in the fields of architecture, communications and education. Bibliometric research methods were applied to rank Australian journals by the JDF, and this was then compared with ranking by JIF. The findings revealed that, JDF rank, when examined alongside total journal articles cited, supported the notion that some journals are highly cited by a relatively small set of journals. That is, the journal's diffusion is limited but may be important to a set of specialized journals. A high JDF was not associated with a journal having an international publisher, nor being indexed by ISI. When the overall ranking of journals was compared, it was found that listing by JIF was positively associated with the overall rank by JDF. The results support Rowland's findings that JDF calculation appears to rank journals in a very different order compared to those listed by JIF. The closest association found in the study was the positive correlation between journal rank for percentage of articles cited and total citations. Similarly, a positive association was found between the numbers of articles cited and total citations. The author argued that JDF ranking is a useful tool when assessing journals for degree of specialization and scholarliness (those with a lower ranked JDF) and degree of trans-disciplinary reception (journals with a higher ranked JDF). The author cautioned that JDF does not provide a comparable alternative journal ranking method to JIF. The JDF may contribute to our understanding of the nature of a journal, but until further research is conducted a JDF ranking should be considered as an independent measure of journal rank.

Haddow and Genoni (2009) discussed the practice and issues related to citation studies and journal ranking in order to provide a suitable indicator for Australian researchers. To achieve this, they examined a sample of Australian education journals and made comparisons three citation-based measures; an extended impact factor, the h-index and diffusion measures, using data drawn from *Scopus* and *Web of Science*. The researchers noted that examining the diffusion of Australian education journals provided an indication of the exposure they gain and may, if other indicators are unavailable, proved to be a useful method to assess a journal as a potential publishing channel. They then used a "six year publication period" to calculate the diffusion measures. Results show that the citation source is a major influence on the total number of citations, the results from *Publish or Perish* (drawing citations from *Google Scholar*) were consistently, and considerably, higher for all journals with the exception of *Higher Education Research & Development*. They noted that without extensive and labor-intensive analysis it was not possible to indicate the degree of overlap between *Scopus* and *WoS*. The study acknowledges the importance of *Scopus* as a citation source for education titles. They suggested that Australian research managers need to continue to develop a journal assessment method that is sensitive to their disciplines.

## OBJECTIVES AND METHOD

The purpose of this paper is to show whether RDF and JDF are good indicators for measuring the breadth of a single journal's influence, as they spread across the literatures through time and to find out if there is any correlation between JDF and JIF.

This study applied publication – citation based assessments methods to present the result of a calculated "equivalent" Impact Factor, Relative Diffusion Factor, and Journal Diffusion Factor for the *Medical Journal of Malaysia* (*MJM*) in both their synchronous and diachronous versions. The publication data was collected from an open access database, Malaysian Abstracting and Indexing system *(MyAis - http://myais.fsktm.um.edu.my)*, while citation data was obtained from *Google Scholar* sourced from the Publish or Perish software (http://wwwxliarzing.com/resources.htm).

 Aware of the issues and shortcomings related to using *Google Scholar* as a data source, the process of data collections followed a very carefully planned step by step approach. The first step is extracting data from the two databases and arranging them according to a specified set of field in each of the spreadsheet accommodating the data. This is followed by thorough data clean up to eliminate errors due to duplication of citations, incorrectly attributed citations, misspellings, incomplete or wrong addresses, publication year and incorrect citations. This study employed the approach encouraged and described by Ingwersen et al. (2001); Frandsen (2004); Frandsen, Rousseau and Rowlands (2006) in calculating the diffusion factors. We created a publication-citation matrix table for *MJM* containing the number of articles published per year and the number of citations received per year from journal articles. We further created an augmented publication – citation matrix table for *MJM* (both synchronous and diachronous versions), to indicate the number of unique new journals involved in citations per year. These tables were used to calculate the impact factor, journal diffusion factor and relative diffusion factor for *MJM* in both their synchronous and dichronous versions.

## FINDINGS

### (a) Publication – Citation Matrix

Citations and impact are always calculated with respect to a certain "pool of citing articles" (Ingwersen et al 2001), in the context of this paper, thepool of articles citing a single journal, *Medical Journal of Malaysia* (*MJM*) within a specified publication window (2004 - 2008) and citation window (2004 - 2010). The publication-citation matrix (in short: p-c matrix) is a handy tool for explaining many notions used in citation analyses (Liang and Rousseau 2009). Thus, the first task is to construct a publication-citation matrix for *MJM* represented in Table 1. The set of data used for this analysis are based on 2004 – 2008 publication year and 2004 – 2010 citation years.

The cells contain the number of articles published per year and the number of citations received per year from journal articles only. The first rows of the table provide the number of published articles per year while the subsequent rows provide the number of citations.

For example, in 2007 *MJM* received 86 citations to the articles published in 2004. In the same year *MJM* received 55 citations to articles published in 2005.

Table 1: Publication – Citation Matrix for *Medical Journal of Malaysia*

| Publication year | 2004 | 2005 | 2006 | 2007 | 2008 |
|---|---|---|---|---|---|
| Number of publications | 139 | 102 | 104 | 100 | 135 |
| Number of citations received from journals in year 2004 | 8 | | | | |
| Number of citations received from journals in year 2005 | 37 | 2 | | | |
| Number of citations received from journals in year 2006 | 76 | 26 | 7 | | |
| Number of citations received from journals in year 2007 | 86 | 55 | 58 | 1 | |
| Number of citations received from journals in year 2008 | 89 | 69 | 60 | 14 | 7 |
| Number of citations received from journals in year 2009 | 69 | 67 | 87 | 52 | 35 |
| Number of citations received from journals in year 2010 | 44 | 30 | 41 | 32 | 30 |

## (b) Impact Factor

IF is calculated by taking the number of all citations for a particular journal for the 2 previous years, and dividing this by the total number of articles published in the journal during that time (Garfield 1993). To calculate the ISI equivalent impact factor of *MJM* for year 2009, we base our calculation on Table 1 with *Equation 1* as follows:

$$IF_2(2009) = \frac{52+35}{100+135} = 0.37 \tag{1}$$

This approach was employed, because ISI impact factor is calculated after 2 years of publications, based on citations received from journal articles alone. Table 1 also represented citations received only from journal articles. Although ISI impact factor is based only on the journals indexed by ISI whereas citations received by *MJM* from non-ISI journals were also included in our calculation.

## (i) Synchronous Impact Factor

The *n*-year "synchronous impact factor" of a journal *J* in the year *Y* is defined by *Equation 2* (Ingwersen et al. 2001; Frandsen, Rousseau and Rowlands 2006; Liang, Rousseau and Egghe 2009). The synchronous impact factor can be calculated for a particular journal, in a particular year, using a particular time window.

$$IF_n(Y) = \frac{\sum_{i=1}^{n} CIT(Y, Y-i)}{\sum_{i=1}^{n} PUB(Y-i)} \tag{2}$$

$CIT_J(Y, X)$ denotes the number of citations received from all members of the pool by a fixed journal *J* in the year *Y*, by articles published in the year *X*. Likewise, $PUB_J(Z)$ denotes the number of articles published by this same journal in the year *Z*. Citation data for a synchronous impact factor will always be found in the same row of the publication-citation matrix.

So, according to Ingwersen et al. (2001); Frandsen, Rousseau and Rowlands (2006); Liang, Rousseau and Egghe (2009), $IF_2$ *(Equation 1)* is a synchronous impact factor involving a single citation year and two publication years. The term "synchronous" refers to the fact that the citations used for the calculations were all received in the same year. In other words, they may be found in the reference lists published in the same year.

Likewise, a 3 year "synchronous impact factor" of *MJM* for the year 2009 is found using *Equation 3*.

$$IF_3(2009) = \frac{87+52+35}{104+108+135} = 0.51 \tag{3}$$

So *Equation 3 is a* three - year "synchronous impact factor" (*IF₃*) for year 2009. The numerator represent the citations in 2009 to articles published in 2006, 2007 and 2008, while the denominator is the number of articles published in 2006, 2007 and 2008, as in Table 1.

The result of our equivalent IF is very close to the results obtained by (Sanni and Zainab 2010) when calculating the equivalent IF of *MJM*, using citation data retrieved from *Google Scholar*. In that study, Sanni and Zainab (2010) considered other citations received from books and other sources. The values were very close due to the fact that *MJM* received 93% of its citations from journal articles.

The "equivalent Garfield IF" (Garfield 1993) and the "synchronous IF" could not be calculated for year 2004, 2005 and 2010 (Table 4) because the "Garfield and Synchronous IF" can only be calculated after 2 years of publication. Note that this present study data samples were articles published between years 2004 – 2008 and citations received between years 2004 – 2010.

**(ii) Diachronous Impact Factor**

The *n*-year "diachronous IF" of a journal *J* in the year *Y* is defined as: *Equation 4* (Ingwersen et al 2001; Frandsen, Rousseau and Rowlands, 2006). The "diachronous Impact Factor" can be calculated for a particular journal, in a particular year, using a particular time window.

$$IMP^s{}_n(Y) = \frac{\sum_{i=s}^{s+n-1} CIT(Y+i,Y)}{PUB(Y)} \tag{4}$$

Where *s*=0, 1, 2, ... denotes a possible shift with respect to the year of publication.

The term "diachronous" refers to the fact that the data used to calculate it are derived from a number of different years with a starting point somewhere in the past and encompassing subsequent years. Thus the "diachronous IF" can be said to reflect the development over time. Citation data for the "diachronous IF" are always found in the same column of the publication-citation matrix.

Therefore, the 2006, two - year "diachronous IFr" for *MJM* is: *Equation 5,* as in table 1, and if we consider the year of publication is: *Equations 6.*

$$IMP_2(2006) = \frac{58+60}{104} = 1.13 \qquad (5)$$

Or, if one includes the year of publication: *Equation 4*

$$IMP^0{}_2(2006) = \frac{7\,|\,50}{104} = 0.62 \qquad (6)$$

Since the "diachronous IF" is influenced by a fixed publication year, then we are able to calculate the "diachronous IF" for each publication year (result in Table 4).

**(d) Journal Diffusion Factors**

*(i) Synchronous Journal Diffusion Factor*
The *n*-year "synchronous journal diffusion factor" for a journal *J* in the year *Y* is: *Equation 7* (Frandsen 2004). The "synchronous journal diffusion factor" can be calculated for a particular journal, in a particular year, using a particular time window.

$$DIF_n(Y) = \frac{\sum_{j=0}^{n-1} U(Y, Y-j)}{\sum_{j=0}^{n-1} PUB(Y-j)} \qquad (7)$$

*U(Y,Y–j)* denotes the number of unique new journals for citations in the year *Y*, to articles published in this journal in the year *Y–j*. The phrase "unique, new" refers to the fact that this journal has not cited an article published in the journal *J* in the years *Y*, …, *Y–*j+1, but that it did cite (in the year *Y*) an article published in the year *Y–j*.

So, to calculate the "synchronous JDF" for *MJM,* we have extended the publication-citation-matrix (Table 1) by including the number of unique new journals that yield the citations (Table 2). In this context, the word "new" refers to the fixed citation year we are considering, and it means that we would consider the matrix row by row and add new journals from the right (the citation year) to the left. This approach leads to Table 2.

For example, *MJM* received 55 citations in 2007 to articles it publish in 2005. These 55 citations occurred in different journals, 41 of which were not involved in a citation (in year 2007) to articles published 2007 or 2006 in *MJM.* It should however, be noted that the total number of different journals involved in those 55 citations were not mentioned, since we do not need to consider this information.

Hence, the 3-year "synchronous JDF" factor of *MJM* for the year 2006 is: *Equation 8*, as in Table 2, it should be noted that any journal can only contribute to the numerator once.

$$DIF_3\ (2006) = \frac{4+18+61}{104+102+139} = 0.24 \qquad (8)$$

The "synchronous JDF" is influenced by a fixed citation year. It means in practice that all the citation window years will have a value if citations occurred every year (result in Table 4).

Table 2: Augmented Publication – Citation Matrix for *MJM* (synchronous version)

| Publication year | 2004 | 2005 | 2006 | 2007 | 2008 |
|---|---|---|---|---|---|
| A | 139 | 102 | 104 | 100 | 135 |
| B (2004) | 8 - 8 | | | | |
| C (2005) | 37 - 33 | 2 - 1 | | | |
| D (2006) | 76 - 61 | 26 - 18 | 7- 4 | | |
| E (2007) | 86 - 66 | 55 - 41 | 58 - 39 | 1 - 1 | |
| F (2008) | 89 - 55 | 69 - 50 | 60 - 41 | 14 - 11 | 7 - 2 |
| G (2009) | 69 - 64 | 67 - 54 | 87 - 77 | 52 - 38 | 35 - 29 |
| H (2010) | 44 - 38 | 30 - 25 | 41 - 35 | 32 - 23 | 30 - 25 |

Notes: A: No of publication per year; B: Number of citations received in year 2004 – number of unique journals involved; C: Number of citations received in year 2005 – number of unique journals involved (new, with respect to the year 2005); D: Number of citations received in year 2006 – number of unique journals involved (new, with respect to the year 2006); E: Number of citations received in year 2007 - number of unique journals involved (new, with respect to the year 2007); F: Number of citations received in year 2008 - number of unique journals involved (new, with respect to the year 2008); G: Number of citations received in year 2009 - number of unique journals involved (new, with respect to the year 2009); H: Number of citations received in year 2010 - number of unique journals involved (new, with respect to the year 2010).

### (ii) Diachronous Journal Diffusion Factor

The *n*-year "diachronous JDF" for a journal *J* in the year *Y* is: *Equation 9* (Frandsen 2004). The diachronous JDF can be calculated for a particular journal, in a particular year, using a particular time window.

$$DI_n(Y) = \frac{\sum_{j=0}^{n-1} U(Y+j, Y)}{PUB(Y)} \qquad (9)$$

$U(Y+j,Y)$ denoted the number of unique new journals involved in citations in the year $Y+j$, to articles published in this journal in the fixed year $Y$. The phrase "unique new" refers to the fact that this journal has not cited an article published in the journal $J$ in the year $Y$ during the years $Y, Y+1, \ldots, Y+j-1$, but that it did cite (in the year $Y+j$) an article published in the year $Y$. It should be noted that the meaning of "unique new" is different here from the "synchronous JDF":

So, to calculate the "diachronous JDF" for *MJM,* we have extended the publication-citation-matrix (Table 1) by including the number of unique new journals that yield the citations (Table 3).

In this context, "new" refers to the fixed publication year that we are considering and it means that we will consider the matrix column by column and add new journals from the top (the publication year) to the bottom which results in Table 3.

Table 3: Augmented Publication – Citation Matrix for *MJM* (diachronous version)

| Publication year | 2004 | 2005 | 2006 | 2007 | 2008 |
|---|---|---|---|---|---|
| **A** | **139** | **102** | **104** | **100** | **135** |
| **B (2004)** | 8 - 7 | | | | |
| **C (2005)** | 37 - 32 | 2 - 1 | | | |
| **D (2006)** | 76 - 56 | 26 - 21 | 7 - 6 | | |
| **E (2007)** | 86 - 55 | 55 - 43 | 58 - 44 | 1 - 1 | |
| **F (2008)** | 89 - 43 | 69 - 52 | 60- 46 | 14 - 13 | 7 - 3 |
| **G (2009)** | 69 - 51 | 67 - 46 | 87 - 74 | 52 - 42 | 35 - 32 |
| **H (2010)** | 44 - 11 | 30 - 21 | 41 - 36 | 32 - 24 | 30 - 28 |

Notes: A: Number of publication per year; B: No of citations received in the year 2004 – number of unique journals involved; C: Number of citations received in the year 2005 - number of unique, new journals involved (new, for the publication year on top of the column, and with respect to previous rows); D: Number of citations received in the year 2006 - number of unique, new journals involved; E: Number of citations received in the year 2007 - number of unique, new journals involved; F: Number of citations received in the year 2008 - number of unique, new journals involved; G: Number of citations received in the year 2009 - number of unique, new journals involved; H: Number of citations received in the year 2010 - number of unique, new journals involved.

For example, *MJM* receives 55 citations in 2007 to articles that it published in 2005. These 55 citations occurred in different journals, 43 of which were not yet involved in a citation (in 2005 or 2006) to articles published in *MJM* in 2005.

Hence, the 3-year "diachronous JDF" (DI) for *MJM* for the year 2006 is defined as: *Equation 10*, as in Table 3.

$$DI_3(2006) = \frac{6+14+46}{104} = 0.92 \qquad (10)$$

In essence, there is no reason to want to limit our calculations to two or three years if we have data available for succeeding years. Set of data available in Table 3 allow us to calculate a seven – years "diachronous JDF" for year 2004, six – years "diachronous JDF" for year 2006 and so forth. The result generated using this approach is represented in Table 4.

## (e) Relative Diffusion factor

### (i) Synchronous Relative Diffusion Factor

The *n*-year "synchronous relative diffusion factor (RDF)" of a journal *J* in the year *Y* is: *Equation 11* (Frandsen et al 2006). The "synchronous RDF" can be calculated for a particular journal, in a particular year, using a particular time window.

$$RDIF_n(Y) = \frac{\sum_{j=0}^{n-1} U(Y, Y-j)}{\sum_{j=0}^{n-1} CIT(Y, Y-j)} \qquad (11)$$

Note that the definition of the relative diffusion factor is different from the journal diffusion factor because rather than dividing by the number of publications, alternatively we are dividing by the number of citations. For example, the 3-year "synchronous RDF" of *MJM* for the year 2006 is: *Equation 12*, as in Table 2.

$$RDIF_3(2006) = \frac{4+18+61}{7+26+76} = 0.76 \qquad (12)$$

### (ii) Diachronous Relative Diffusion Factor

The *n*-year "diachronous relative diffusion factor (RDF)" of a journal *J* in the year *Y* is defined as: *Equation 13* (Frandsen, Rousseau and Rowlands 2006). The diachronous RDF can be calculated for a particular journal, in a particular year, using a particular time window.

$$RDI_n(Y) = \frac{\sum_{j=0}^{n-1} U(Y+j, Y)}{\sum_{j=0}^{n-1} CIT(Y+j, Y)} \qquad (13)$$

For example, 5-year "diachronous RDF" for *MJM* for the year 2006 is defined as: *Equation 14*, as in Table 3.

$$RDI_5(2006) = \frac{6+44+46+74+36}{7+58+60+87+41} = 0.814 \qquad (14)$$

It should be noted also that any journal can only contribute to the numerator once.

Table 4 shows the result of the calculated "diachronous relative factor". The year 2009 and 2010 rows have no values because the calculation of a "diachronous relative factor" is influenced by the fixed publication year(s) we are considering. Practically, for year 2004, we can calculate a 7- year- "diachronous relative factor" for *MJM*. Thus, $RDI_7$ (2004) is a diachronous (Relative Diffusion Factor) involving a single publication year and 7 citation years. Likewise, $RDI_6$ (2005) is a "diachronous relative factor" involving a single publication year and 6 citation years.

$$RDI_7(2004) = \frac{7+32+56+55+43+51+11}{8+37+76+86+89+69+44} \tag{15}$$

$$RDI_6(2005) = \frac{1+21+43+52+46+21}{2+26+55+69+67+30} \tag{16}$$

As argued under the explanation of the "diachronous JDF", we believe there is no basis for us to constrain the calculations to two or three years if we have data available for succeeding years. In addition, our results has confirmed that "the larger the number of citations, the lower the *RDI* becomes, suggesting a negative correlations" (Frandsen, Rousseau and Rowlands 2006). The year 2004 recorded the highest number of citations (409 citations), and has gotten the lowest RDI (0.62). While year 2008 recorded the lowest number of citations (72), and have the highest RDI (0.88).

This is so because, the numerator (the number of different citing journals) of the fraction cannot vary as much as the denominator (the number of citations) and therefore the highly cited journals/publication-period will be in a situation of low RDI, as any extra citations from an already citing journal will reduce the RDI (Frandsen, Rousseau and Rowlands 2006).

Table 4: Synchronous and Diachronous IF, RDF, and JDF for *MJM*

| Year | Impact Factor | | | Relative Diffusion Factor | | Journal Diffusion Factor | |
|------|------|------|------|------|------|------|------|
| | Equivalent | Synchronous | Diachronous | Synchronous | Diachronous | Synchronous | Diachronous |
| | Garfield impact factor | $(IF_2)$ | $(IMP_2)$ | $(RDIF_n(Y))$ | $(RDI_n(Y))$ | $(DIF_n)$ | $(DI_n)$ |
| 2004 | x | x | 0.81 | 1 | 0.62 | 0.057 | 1.83 |
| 2005 | x | x | 0.79 | 0.87 | 0.74 | 0.141 | 1.8 |
| 2006 | 0.42 | 0.42 | 1.13 | 0.77 | 0.81 | 0.24 | 1.98 |
| 2007 | 0.55 | 0.55 | 0.66 | 0.74 | 0.81 | 0.33 | 0.8 |
| 2008 | 0.36 | 0.36 | 0.48 | 0.66 | 0.88 | 0.27 | 0.46 |
| 2009 | 0.37 | 0.37 | x | 0.85 | x | 0.45 | x |
| 2010 | x | x | x | 0.82 | x | 0.25 | x |

## LIMITATIONS

This study is limited by the fixed publication and citation years examined and affected by any shortcomings of the databases (*MyAis and Google Scholar*) from which data were retrieved. Analysis of citation data were also limited to journal articles while citations that fall into the categories grey literature such as unpublished reports and papers, preprints, PowerPoint presentations, theses, and conference papers were omitted.

## CONCLUSIONS AND DISCUSSIONS

In this study, we have been able to present the results of our calculated "equivalent" Impact Factor, Relative Diffusion Factor, and Journal Diffusion Factor for the *Medical Journal of Malaysia* (*MJM*) in both their synchronous and diachronous versions. The values of the synchronous journal diffusion factor ranges from 0.057 – 0.14 while for the diachronous journal diffusion factor is in the range 0.46 – 1.98. The high diachronous journal diffusion factor recorded by the articles published in year 2006 ($DI_n$ = 1.98) can be explained by a relatively high number of different citing journal (206) against the number of publications (104). Publications in year 2004 (139) have the highest number of different citing journals (255), but have a diachronous journal diffusion factor value ($DI_n$ = 1.83) lesser that year 2004 because of a considerable high numbers of publication. This implies that the results of diachronous journal diffusion factor is influenced by the numbers of publications, which means that a good comparison may be one of which the subject of analysis have the same or similar number of publications and citations period. The yearly values of the synchronous relative diffusion factor vary in the range 0.66 – 1.00 while diachronous relative diffusion factor ranges from 0.62 – 0.88. The result shows that diachronous relative diffusion factor has negative correlations with the number of citations, resulting in a low *RDI* value for highly cited publication years; thus, the way to go is to compare like with like in journal evaluations. Results from this study also suggest that the Journal Diffusion Factors should be considered as an independent measure of journal rank as there is no particular pattern when the results of the different JDF are compared. Nonetheless, one significant outcome of the diachronic version of the Relative Diffusion Factor and Journal Diffusion Factor is that they can be applied at all levels of evaluations. They can be applied to a single article, to a series of articles published in a specific issue or a particular year. When applied, it can indicate the continuous influence of an article or series of articles across time in different literatures. What this implies in practice is that this diffusion factors can be calculated every additional year at any journal level of analysis. Thus, this study has been able to demonstrate that these indicators are valuable tools that help to show development as it changes through time. Findings also suggest that the diachronous version of the diffusion factors will be of much interest to policy makers, editorial committees of local or regional journals who often have a research focus for a particular year or economic period and may wish to compare the influence of a research flow of a period against another, which in effect can help to understand the focus, and identify strategies employed that results in the spread of the unit studied across bodies of literatures through time.

In conclusion, our findings shows that JDF and RDF are good indicators for measuring the breadth of influence of a single journal within a certain period of time because it captures different aspects of the citation reception process. These indicators can definitely be used alongside the synchronous and diachronous impact factor at any level of journal assessment.